# Coherent Frequency Combs produced by Self Frequency Modulation in Quantum Cascade Lasers

J. B. Khurgin[1], Y. Dikmelik[1], A. Hugi[2], J. Faist[2]

[1]Department of Electrical and Computer Engineering, Johns Hopkins University, Baltimore, Maryland 21218, USA
[2]Institute for Quantum Electronics, ETH Zurich, Wolfgang-Pauli-Strasse 16, 8093 Zurich, Switzerland

One salient characteristic of Quantum Cascade Laser (QCL) is its very short $\tau \sim 1ps$ gain recovery time that so far thwarted the attempts to achieve self-mode locking of the device into a train of single pulses. We show theoretically that four wave mixing, combined with the short gain recovery time causes QCL to operate in the self-frequency-modulated regime characterized by a constant power in time domain and stable coherent comb in the frequency domain. Coherent frequency comb may enable many potential applications of QCL's in sensing and measurement.

Quantum cascade lasers (QCL's)[1] are versatile sources of coherent radiation in the mid and far infrared regions of the spectrum, in which many chemical compounds have characteristic strong fundamental absorption lines. The atomic-like joint density of state character of intersubband transitions, combined with quantum engineering[2-4], allows the development of QCLs with heterogenous active regions[5] covering a very large spectral range. Recently, by combining such broadband active regions with an external cavity[6], continuous tuning over large spectral ranges have been achieved in the 8-10 and 3-5μm wavelength range[7]. However, these sources, tunable over hundreds of wavenumbers, require a cavity with moving components, naturally limiting their tuning speed as well as their compactness.

In the last decade, a new approach to broadband spectroscopy has been proposed and implemented by the use optical frequency combs[8, 9]. In a dual comb spectrometer[10-12], two optical combs with slightly different mode spacing are beaten onto a fast detector. Since each pair of modes (one from each combs) can be made to beat at a slightly different frequency, a Fourier transform of the signal on the detector enables a direct reading of the optical spectrum. So far, such technique has been demonstrated using combs based on mode-locked lasers, in which the periodic time optical output, consisting of pulses spaced by the round-trip period of the cavity, ensures the periodicity of the comb optical spectrum.

To reduce the power consumption and size, it would be very attractive to generalize this technique for semiconductor lasers. Furthermore, for many applications, it would be especially interesting to do so in the mid-infrared region of the spectrum, where many molecules have their strong fundamental absorption lines and where the generation of combs is more challenging[13]. However, in the conventional free-running multi-mode laser the modes are not evenly spaced due to material dispersion and the broadening of the individual modes prevent the generation of well-separated beat notes in the RF domain. As a result, some way of locking the modes together must be employed. Yet all the numerous attempts[14-16] to mode-lock QCL had limited success, and the practical achievements did not match numerous theoretical predictions. This problem can be traced to the fact that the in the QCL the relaxation time between the upper and lower laser levels (often referred to a "gain recovery time"), $\tau_{21}$ is measured in picoseconds, and is thus much shorter than the cavity round trip time $\tau_{rt}$ that is typically about 50-100ps in a few mm long cavity. The QCL medium thus acts as a fast saturable gain or "reverse saturable absorber". While fast saturable absorber used as mode locker favors short pulses, the fast saturable gain tends to favor CW radiation. While each time the pulse passes through the fast absorber it sharpens, each time the pulse passes through the fast saturable gain of QCL it broadens. Since gain in the laser is by definition stronger than loss the short pulses cannot develop. Even if one could succeed in getting short pulses with active mode-locking this arrangement would be extremely energy inefficient as all the pump current flowing through QCL in between the pulses (i.e. most of the time) would be wasted. Since the energy cannot be efficiently stored in the QCL gain medium between the pulses and then released in a short powerful burst (as in a normal mode-locked laser) the peak power of such actively mode-locked QCL would remain far lower than the average power of CW QCL. Note that QCL operating in the terahertz do not present such strong limitations because of the longer upper state lifetime[17].

At this point it is important to recognize that while the term "mode locking" is most often is used in the narrow sense of the phases of all modes being equal; in the wider, and more proper sense, the term "mode-locking" only implies that the phases of all modes are locked in some stable relationship between each other, and this relationship can be far richer than a uniform phase over the whole spectrum. Once this stable relationship is established the electric field of the laser becomes periodic in time and in frequency domain a stable frequency comb arises. This frequency comb can be just as useful as the one produced by short pulses, and, in fact can be superior to the latter as in the absence of high peak power one can avoid deleterious nonlinear effects, optical damage and detector saturation. Recently, frequency comb operation of broadband quantum cascade laser with low group velocity dispersion was reported[18]. In this devices, the experimental characterization of the comb indicated that the relative phases of the modes did roughly correspond to a frequency modulated output with an approximately constant intensity. In this paper we show theoretically how these FM combs are self-generated without resorting to either active modulation or additional passive intra-cavity elements, and we compare these prediction with measurements of the beatnote spectroscopy.

To understand the self-phase-locking mechanism in the QCL let us first consider how the passive model locking is achieved in conventional solid state or semiconductor laser. In time domain the saturable absorber favors short pulses – the peak of the pulse experiences lower loss and thus the pulse sharpens each time it passes through the absorber. In frequency domain four wave mixing (FWM) locks the frequencies and phases of all modes and thus mitigates the effect of dispersion. In QCL the saturable gain favors constant power output as all the peaks tend to be smoothed out by each passage through the gain medium. The most obvious way constant power can be maintained is if the QCL operates in a single mode regime, but the latter mode of operation is quickly destabilized by spatial hole burning[16]. In fact, one has to go to a great length and introduce spectrally selective elements (gratings) into the cavity to attain a single mode operation. Hence on one hand the broad gain and spatial and spectral hole burning in QCL support multimode operation but on the other hand the fast saturable nature of the gain favors constant power. This combination is a signature of a periodic frequency-modulated (FM) signal. FM lasers have been known for a long time and have been successfully operated usually employing an active phase modulator inside the cavity, with the lonely case of passive FM modulation reported in [19]. In contrast, in QCL the FM is achieved naturally by the process of FWM in the gain medium itself[20] which despite presence of a certain amount of dispersion locks the frequencies of the modes in such way that they are separated by $\tau_{rt}^{-1}$ and the phases of the modes in such way that all the intermodal beat oscillations cancel and the power of the QCL is almost constant.

To understand how the self-FM modulation arises we have developed a very simple theoretical description in which the evolution of the amplitude of n-th mode is described by one in the set of N equations

$$\frac{dA_n}{dt} = (G_n - 1)A_n - G_n \sum_{k,l=-N/2}^{N/2} A_m A_k A_l^* B_{kl} C_{kl} \kappa_{klmn} \qquad (1)$$

where $m=n-k+l$, the time has been normalized to the photon lifetime, gain in the $n$-th mode $G_n = G_0 \left[1 + n^2 \left(\tau_{coh}/\tau_{rt}\right)^2\right]^{-1}$ where $G_0$ is the unsaturated gain of the mode resonant with the intersubband transition (normalized to the threshold), the coefficients $B_{kl} = \left[1 + \tfrac{1}{2}(k^2 + l^2)(\tau_{coh}/\tau_{rt})^2\right]^{-1}$ where $\tau_{coh}$ is the coherence time of the transition determines how many modes participate in FWM and coefficients $C_{kl} = \left[1 - j(k-l)\tau_{21}/\tau_{rt}\right]^{-1}$ describe the amplitude of coherent population beating, while $\kappa_{klmn} = l_c \int_0^{l_c} \sin(k_k x)\sin(k_l x)\sin(k_m x)\sin(k_n x)dx$ is the normalized intermodal overlap in space. The model contains a number of simplifications, such as it is perturbative to the third order and, due to limitations of computer power uses number of modes (less 100) that is less than in a typical QCL. Nevertheless, our limited goal here is not to model each and every operating characteristics of QCL, but just to convincingly demonstrate that FM frequency comb is a natural operating regime of QCL and thus clarify theoretically the origin of the frequency combs observed in[18]. In addition, we assumed that the broad gain bandwidth of QCL combs as well as operation in the 6-8µm range allowed us to neglect all dispersion including the one by the lasing transition itself (Ref 18).

Consider now special cases of the frequency mixing terms in (1). First of all, there is a fully degenerate term $k=l=m=n$ that describes self-saturation of gain mode $n$. Then there are N degenerate terms $k=l$, $m=n$ that describe the cross-saturation of the gain in mode $n$ by the intensity in mode k. For all these terms the coefficient $C_{kk}=1$. Furthermore, since for the self saturation the modal overlap $\kappa_{nnnn}=3/8$ is larger than the one charaterizing cross-saturation $\kappa_{kknn}=1/4$, the multimode operation produces less gain saturation than a single mode operation. Hence it will have lower threshold and the multimode regime will ensue. This is of course nothing but a spectral domain description of spatial hole burning. Then there are $N(N-1)$ non-degenerate terms that describe FWM in which the population beating caused by interference of modes $k$ and $l$ modulates the mode m and creates the sideband in the vicinity of mode $n=m+k-l$. The mode overlap is $\kappa_{klmn}=1/8$. In most known laser media $\tau_{21}>>\tau_{rt}$ and therefore the coefficients $C_{kl}$ are very small indicating that carrier population cannot follow the beating of intensity and the FWM terms are absent leaving the modes uncoupled. In QCL, in contrast the modes are all coupled to each other by FWM causing frequency and phase locking.

To demonstrate how the short gain recovery time causes phase locking we first consider a laser with a long $\tau_{21}\sim$1ns.-on the order of recombination time in a typical diode laser. With all non-degenerate FWM terms eliminated the rate equations become

$$\frac{dA_n}{dt}=(G_n-1)A_n-G_nA_n\left[\tfrac{3}{8}|A_n|^2 B_{nn}+\tfrac{1}{4}\sum_{k\neq n}|A_k|^2 B_{kk}\right]+N_n \qquad (2)$$

where $N_n$ is the spontaneous emission noise. Solving numerically the equations with $G_0=1.4$ and $\tau_{coh}=0.2$ps yields the multi-mode spectrum (Fig.1a) and the strong chaotic modulation of optical power in the time domain (Fig 1b). If one detects such a signal and analyzes its spectrum one should see that all the mode beats are present in the RF spectrum as shown in Fig.1c. From that point of view the chaotic behavior is perfectly reasonable as the gain medium reacts only to the average power inside the cavity hence the chaotic behavior remains unmitigated.

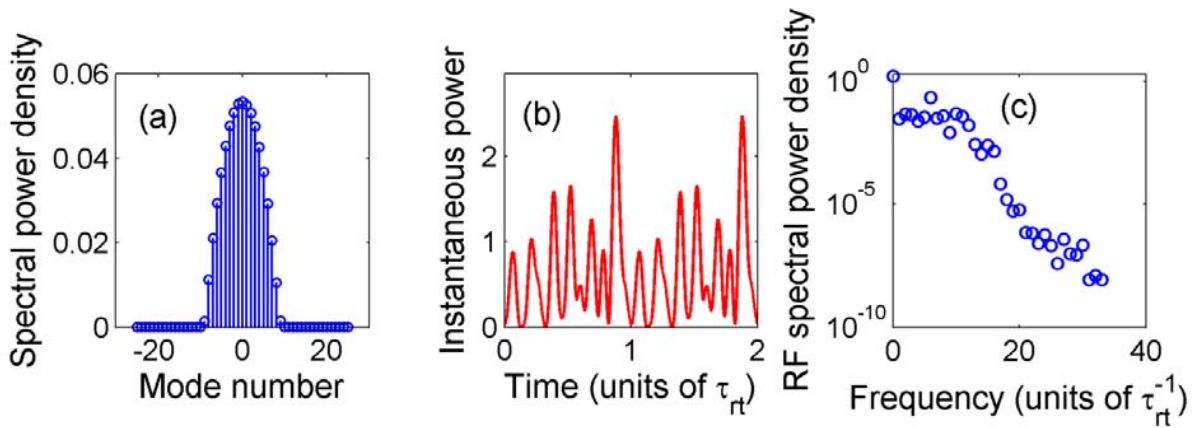

**Fig.1 a.** – Optical Spectrum **b-**Time dependence of Optical Power and **c-** RF Spectrum of detected power of a free-running multi-mode laser with long relaxation time $\tau_{21}=$1ns.

Now, to understand how the mode locking can occur in this laser we add the fast saturable absorber with $\tau_{21(a)}=1.0$ps, $\tau_{coh(a)}=0.1$ps, and unsaturated absorption (loss) $L_0=0.3$ so that the rate equations now become

$$\frac{dA_n}{dt} = (G_n - L_n - 1) A_n - G_n A_n \left[ \frac{3}{8}|A_n|^2 B_{nn} + \frac{1}{4} \sum_{k \neq n} |A_k|^2 B_{kk} \right] + L_n A_n \sum_{k,l} A_m A_k A_l^* B_{kl}^{(a)} C_{kl}^{(a)} \kappa_{klmn}^{(a)} + N_n \quad (3)$$

The results are shown in Fig.2 and include a rather smooth multi-mode spectrum in Fig.2a, a train of perfectly mode-locked pulses in Fig.2b, and a set of well defined beat frequencies at multiple of $\tau_{rt}^{-1}$ in Fig.2c – quite similar to the experimental results indicating that our model is capable of predicting phase locking of the modes.

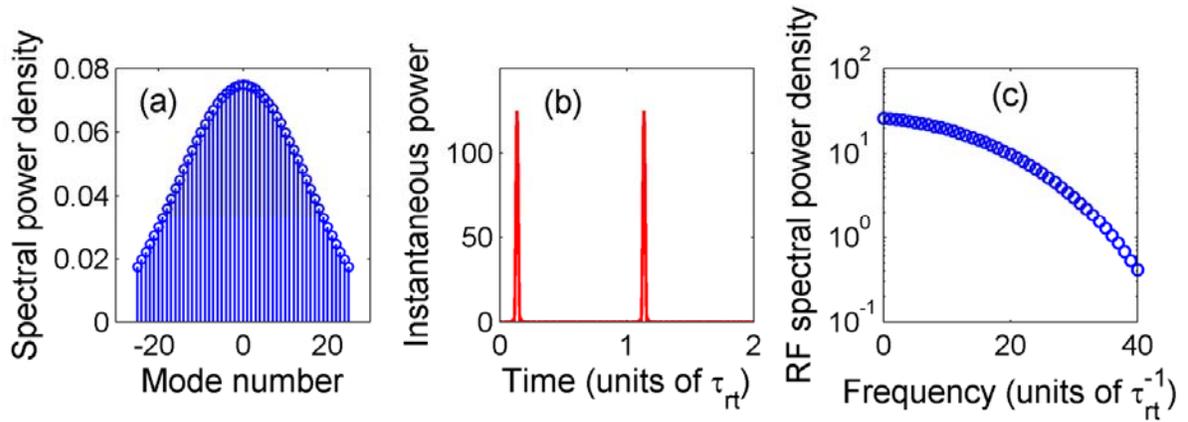

**Fig.2 a.** – Optical Spectrum **b-**Time dependence of Optical Power and **c-** RF Spectrum of detected power of a multi-mode laser with long relaxation time $\tau_{21}=1$ns mode-locked with a saturable absorber

Now we return to the original equation (1) and proceed to solve it for the QCL with $\tau_{21}=1$ps, $\tau_{rt}=65$ps, $\tau_{coh}=0.1$ps and $G_0=1.2$. The results are plotted in Fig.3. The multi-mode spectrum shown in Fig3.a does not differ significantly from the one for a diode laser (Fig.1a), but when it comes to instant power shown in Fig3b the difference is dramatic - the power stays practically constant with a very little amplitude noise. The RF spectrum shown in Fig.3c shows that all the beat frequencies are all suppressed by as much as 30dB, indicating that the phases of the modes are locked by the FWM in the arrangement that cancels all the beat oscillations. In fact, our model predicts a beatnote at the roundtrip frequency amounting to $10^{-3}$ of the c.w. value, close to the value ($\sim 2 \times 10^{-2}$) measured in ref. 18. As we have discussed above, constant power and multi mode operation are compatible only for the case of frequency modulation. The instant frequency is plotted in Fig.3d and one can see that it is strongly modulated. The modulation cannot be described by a nice sinusoidal form but it is clearly periodic, indicating that the modes are coherently locked. This result is of course expected – while there may exist just one mode-locked solution that minimizes the threshold of laser with fast saturable absorber, there are many equivalent FM solutions that minimize the threshold of laser with fast saturable gain. Thus one cannot predict which solution will develop, but what is important is that all these solutions result in perfect frequency comb observed in experiment.

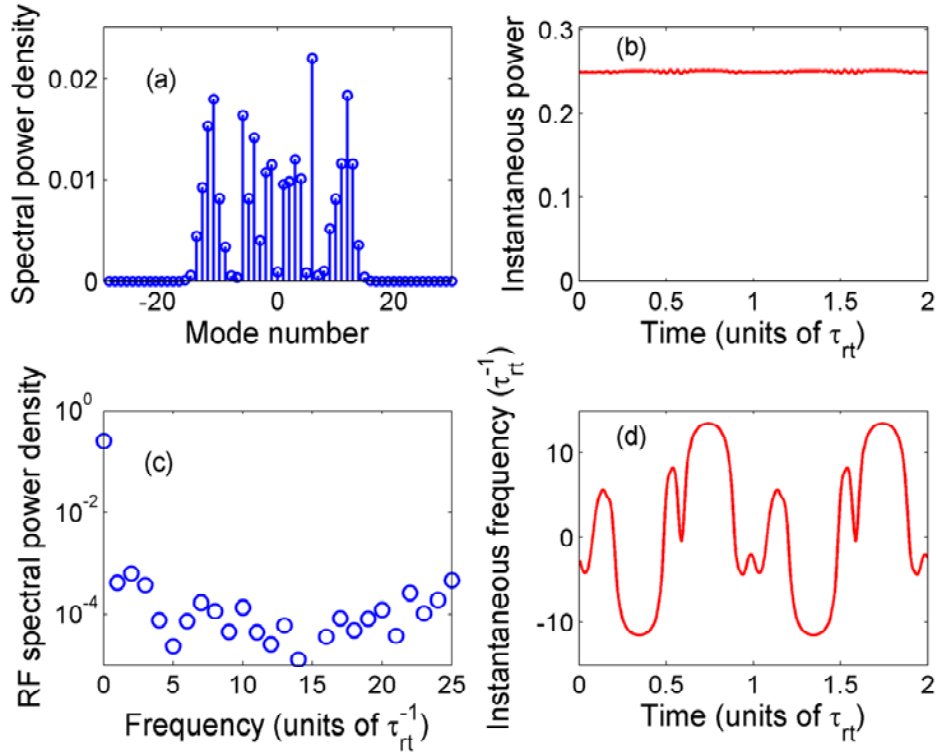

**Fig.3 a.** – Optical Spectrum **b-**Time dependence of Optical Power **c-** RF Spectrum of detected power and **d** –Instant frequency of a free-running multi-mode QCL with a short relaxation time $\tau_{21}$=1ps.

Thus our simple model predicts correctly the general trends of laser behavior – the lasers with long gain recovery time operate in the multi-mode regime and can be mode-locked with a saturable absorber, while QCL with inherently short gain recovery time tends to settle in to the FM regime. In the FM regime the amplitudes and phases of modes are arranged in a delicate balanced relationship, such that the power is constant and the beats of detected power are all cancelled. Clearly any disturbance, such as frequency-dependent loss or group velocity dispersion upsets this delicate balance and frequency modulation (FM) gets converted into the amplitude modulation (AM). This is the standard way of detecting FM signal by using frequency discriminator to convert it into AM signal. Consider for example what happens with the calculated FM output of QCL of Fig.3.a when it passes through the ramp filter with transmission changing linearly with frequency, i.e an optical frequency discriminator. As one can see from Fig.4.a the frequency modulation of Fig.3d is transferred into the amplitude modulation and the detected RF spectrum shows the strong beats at the $\tau_{rt}^{-1}$ ~7.5GHz frequency, shown in

Fig.4b that is increased by 19 dB relative to the ones without discriminator that is also re-plotted in Fig4.b for comparison). This is exactly what has been reported in ref [18]. Fig. 4c shows the measurement of the beatnote before and after the insertion of the discriminator. The QCL in this measurement is run in a comb-like regime (I = 490 mA, T= -20° C) where the perfect locking is lost, as described in ref18. The beatnote has a linewidth of about 100 kHz. Due to the imperfect locking in this operation condition, a noise floor originating from unlocked modes at -107 dBm is visible. Putting the optical discriminator has two effects. First, it increases the intensity of the beatnote formed by the locked modes by 16 dBm. This is expected due to the transfer of frequency modulation into amplitude modulation. However, the intensity of the beating of the unlocked modes forming the noise floor is decreased by 3 dBm. This is also expected, since the discriminator is expected to decrease the signal strength on modes that do not feature a FM phase relationship.

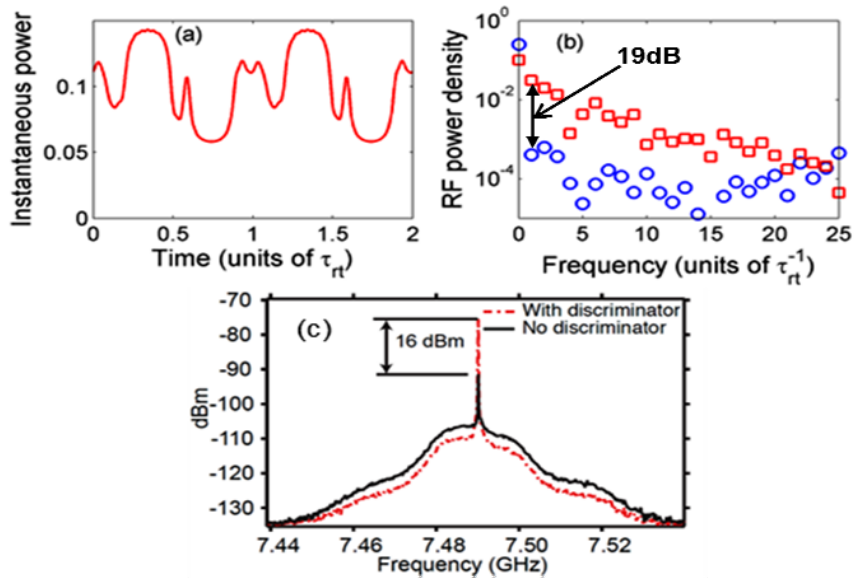

**Fig.4 a.** Time dependence of Optical Power of a free-running multi-mode QCL with a short relaxation time $\tau_{21}$=1ps following a passage through a discriminator **b.** RF spectrum of detected power after discriminator (squares) and before discriminator (circles). **c**. Measurements of RF power spectrum near the beat frequency. The QCL is driven in a comb-like regime at I = 490 mA at a temperature T= -20° C.

Most characterization techniques of mode-locked lasers leverage on the high peak intensities reached during the short pulses, and therefore are of little use for FM optical combs. In ref 18, the beatnote interferogram was measured, in which the RF signal at beat frequency $\tau_{rt}^{-1}$ is measured after passing through a Michelson interferometer (FTIR[21]). As the interferometer periodically becomes unbalanced the dispersion causes the FM-to-AM conversion and RF signal increases and decreases as shown in Fig.5a. This behavior, predicted by our simple model is compared to the result of an actual experiment in Fig. 5b. In the latter, a QCL comb similar to the one used in Ref. 18, is operated in the optical comb regime (I = 475 mA, T = -20 °C) such that the beatnote at (ν = 7.4892 GHz) has a linewidth of < 100Hz is fed into our Brucker 66S infrared spectrometer. The intensity of the beatnote at ν = 7.4892 GHz, recorded by a fast QWIP detector and a RF spectrum analyser as a function of the mirror delay, is then plotted in Fig. 5b and compared to the predictions of the model. There is a very good qualitative agreement between the two curves, in particular the minimum observed for zero path delay that is a characteristics of the FM modulated output, as well as the relative positions and intensities of the first two maxima. In both the simulation and experiment, the second maximum is stronger in intensity. In fact, the measurement shown in Fig. 5b proves that the frequency modulation of the laser is indeed not of a nice sinusoidal form. The first maximum of a beatnote interferogram of such a laser would be stronger by a factor of about 3 compared to the second maximum. Similar characteristics were already observed in Ref. 18.

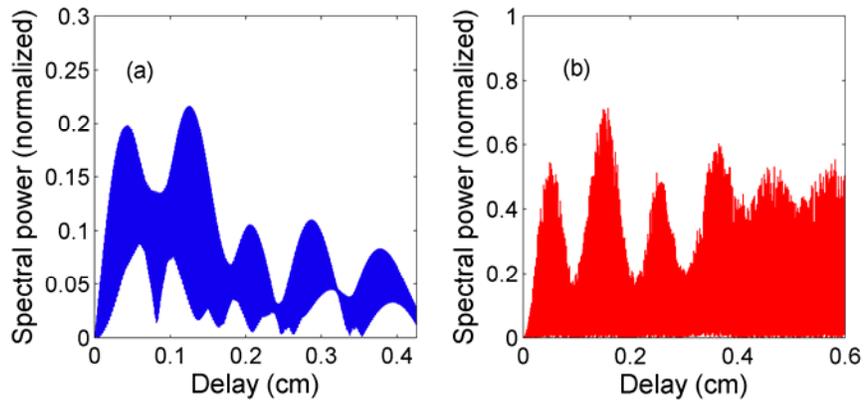

**Fig.5** beat frequency interferogram of QCL **a** Numerical model results **b** Experimental Data - the QCL is operated in the optical comb regime (I = 475 mA, T = -20 °C)

In conclusion, using a simple theoretical model we have demonstrated how short gain recovery time in QCL's almost inevitably leads to a self-frequency modulated operating regime and thus have explained the origin of the experimentally observed coherent frequency combs observed experimentally. The new understanding should facilitate design and implementation of numerous QCL-based instruments for sensing and measurement as well as help the design of QCL combs with broader frequency comb coverage.

**Acknowledgement:** The work at JHU was supported by Mid-Infra-Red Technologies for Health and the Environment Research Center (National Science Foundation Grant MIRTHE NSF ERC; EEC0540832)

**Figure captions:**

**Fig.1 a.** – Optical Spectrum **b-**Time dependence of Optical Power and **c-** RF Spectrum of detected power of a free-running multi-mode laser with long relaxation time $\tau_{21}$=1ns.

**Fig.2 a.** – Optical Spectrum **b-**Time dependence of Optical Power and **c-** RF Spectrum of detected power of a multi-mode laser with long relaxation time $\tau_{21}$=1ns mode-locked with a saturable absorber

**Fig.3 a.** – Optical Spectrum **b-**Time dependence of Optical Power **c-** RF Spectrum of detected power and **d** –Instant frequency of a free-running multi-mode QCL with a short relaxation time $\tau_{21}$=1ps.

**Fig.4 a.**Time dependence of Optical Power of a free-running multi-mode QCL with a short relaxation time $\tau_{21}$=1ps following a passage through a discriminator **b.** RF spectrum of detected power after discriminator (squares) and before discriminator (circles). **c**. Measurements of RF power spectrum near the beat frequency. The QCL is driven in a comb-like regime at I = 490 mA at a temperature T= -20° C.

**Fig.5** beat frequency interferogram of QCL **a** Numerical model results **b** Experimental Data - the QCL is operated in the optical comb regime (I = 475 mA, T = -20 °C)

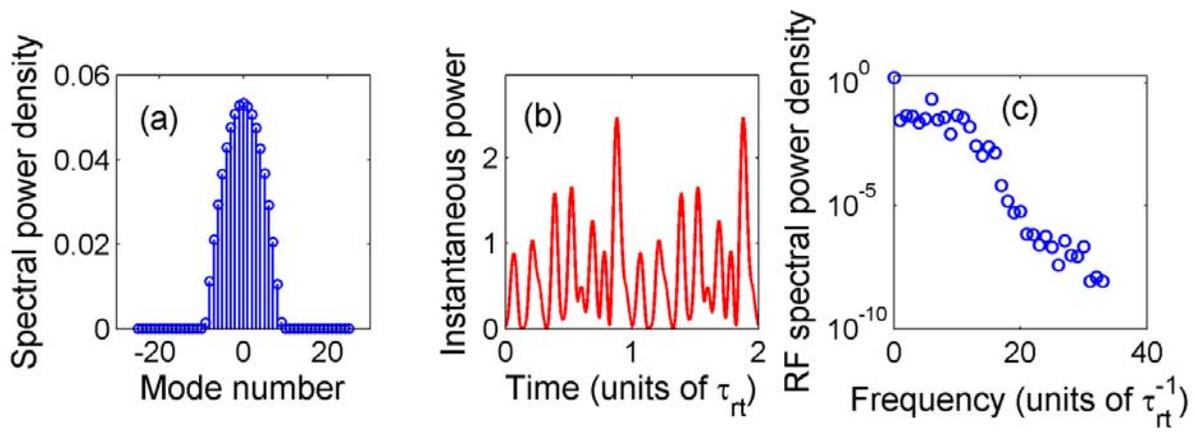

Figure 1

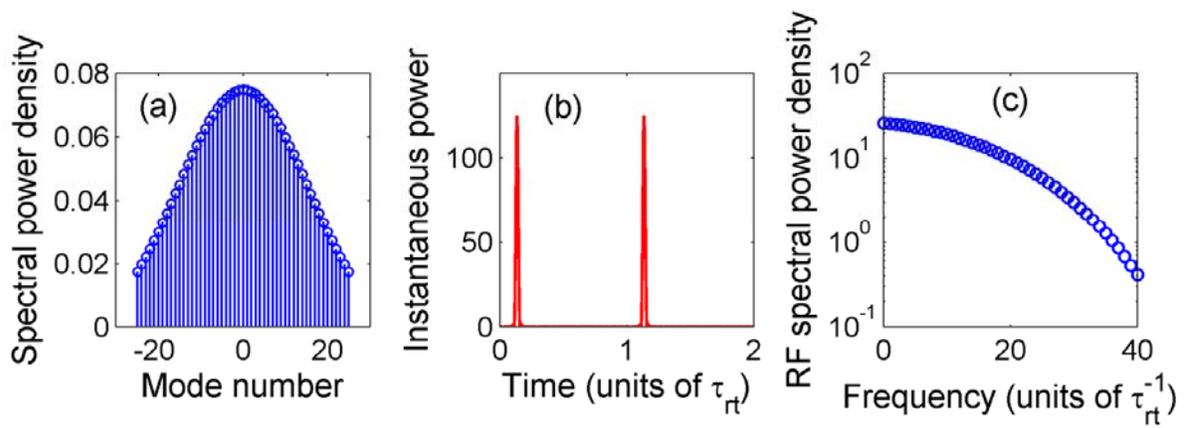

Figure 2

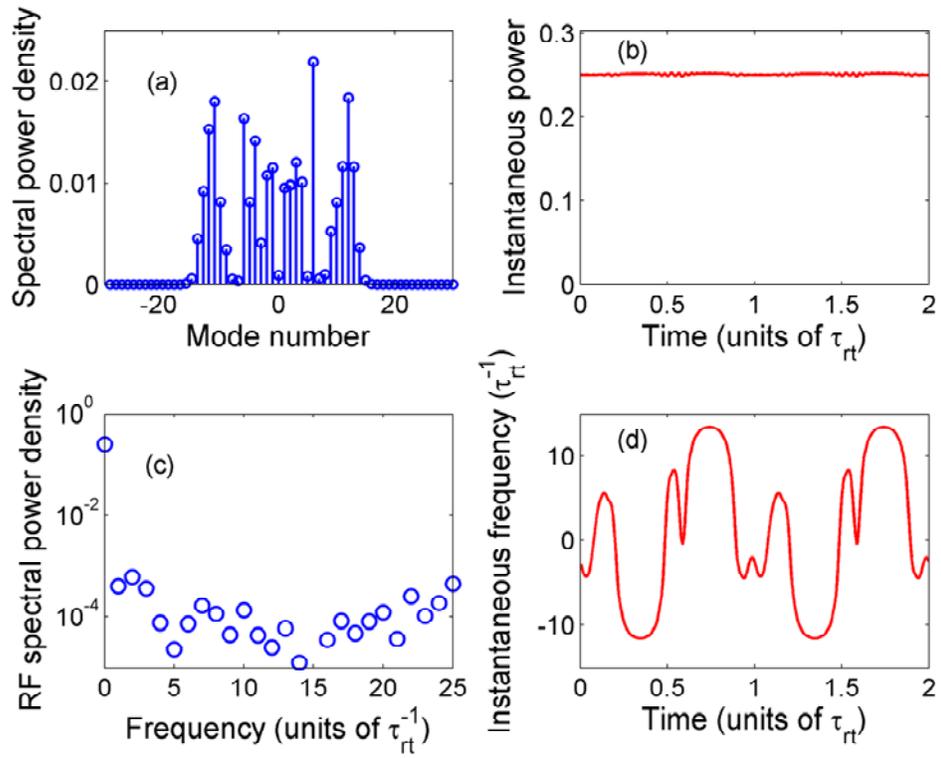

Figure 3

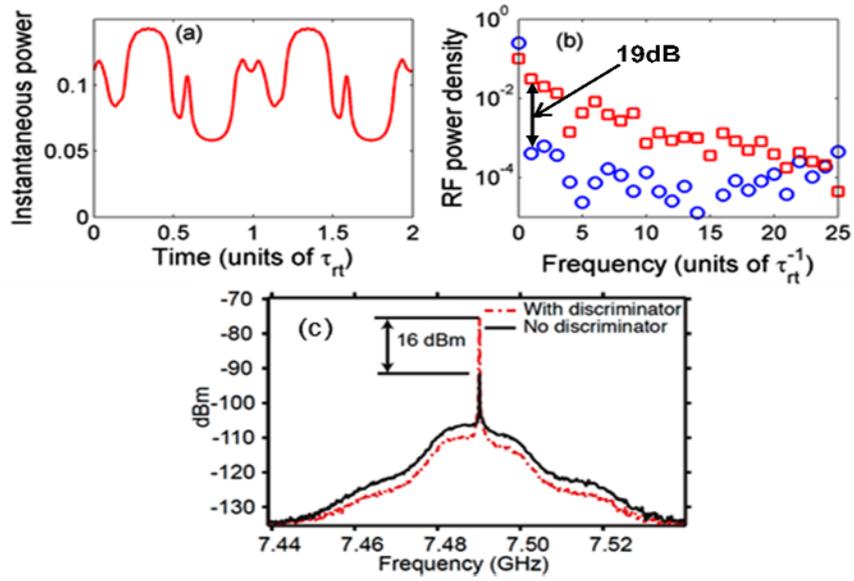

Figure 4

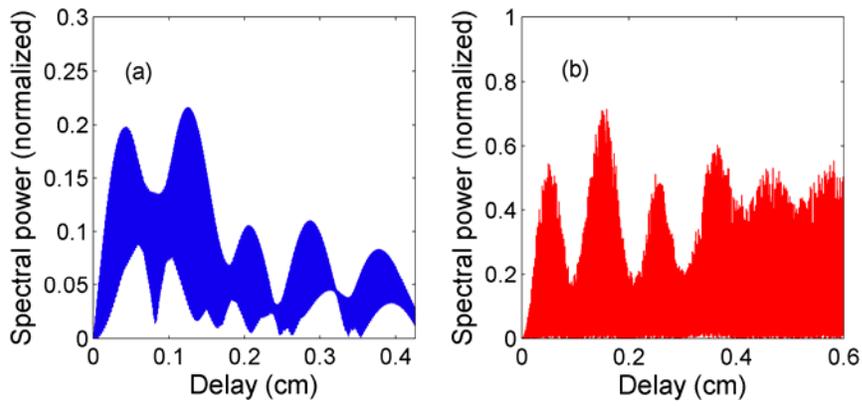

Figure 5